# Why does attention to web articles fall with time?

M.V. Simkin and V.P. Roychowdhury
Department of Electrical Engineering, University of California, Los Angeles, CA 90095-1594

We analyze access statistics of a hundred and fifty blog entries and news articles, for periods of up to three years. Access rate falls as an inverse power of time passed since publication. The power law holds for periods of up to thousand days. The exponents are different for different blogs and are distributed between 0.6 and 3.2. We argue that the decay of attention to a web article is caused by the link to it first dropping down the list of links on the website's front page, and then disappearing from the front page and its subsequent movement further into background. The other proposed explanations that use a decaying with time novelty factor, or some intricate theory of human dynamics cannot explain all of the experimental observations.

## 1. Introduction

Soon after the internet became popular after the development of the first browsers researchers started to seek the laws of web surfing. Cunha, Bestavros, & Crovella (1995) had studied the browsing patterns of the internet users from Boston University. They found that the distribution of web pages by the total number of downloads by all users follows a power law. Nielsen (1997) had observed that the same power law holds for the number of downloads of different webpages from Sun Microsystems website. Huberman et al (1998) looked at the number of webpages downloaded from Georgia Tech website by particular users. They found that the distribution of the number of users by the number of clicks they do follows a power law.

Later researchers got to study the patterns of accessing particular webpages. Dezsö et al (2006) had studied the access log of one news website. They found that web documents are mostly accessed during first days after their creation, with the number of accesses decreasing with time as a power law

$$n(t) \sim 1/t^\beta \qquad (1)$$

with $\beta \approx 0.3$. They explained this using queuing based human dynamics model. Wu and Huberman (2007) have studied the time series of the number of "diggs" (bookmarks) on Digg website. Their work is related to the work of Dezsö et al (2006) because, the number of new bookmarks should be proportional to the number of new accesses. Wu and Huberman found that the number of new "diggs" decreases as the time passed since the story appeared on the website increases. They introduced the concept of a novelty factor, decaying with time as a stretched exponent, to explain their results.

Hogg & Lerman (2009), who also studied Digg, instead of novelty factor, argued for visibility factor. The story first appears on website's front page, as time goes, newly added stories push it down to the page two, page three and so on. The further the story is from the front page the less visible it is, less people notice it, and less the number of new "diggs" it generates. When new stories are added at a constant rate, the position of the story on the website is proportional to the time passed since its publication. The frequency of accessing the story is thus some decreasing function of time. To find the exact functional dependence Hogg & Lerman (2009) devised a theory based on the model introduced by Huberman et al (1998) to describe the distribution of the number of users by the number of pages views. The theory of Hogg & Lerman leads to Eq.(1) with $\beta = 1/2$.

Leskovec et al (2007) studied the evolution of connectivity in blogosphere. They found that the average number of new in-links to a blog entry falls of as an inverse power of time passed since publication of this blog entry. The exponent of the power law is 1.5. We may reason that the number of new in-links is proportional to the number of views (just like the number of bookmarks).

In this article, we study the patterns of access of blog entries and news articles from more than one hundred different websites. We find that for about 80% of them the access frequency decays according to Eq.(1). The exponents are different for different websites and are distributed between 0.6 and 3.2 with the maximum of the distribution around 1.5. We offer a theoretical explanation of these observations. Our explanation is similar to that of Hogg & Lerman (2009) as it is also based on visibility factor. It is different, however in other aspects and allows for different values of the exponent $\beta$ in agreement with our data.

The paper is organized as follows. In Section 2, we describe our data source and analyze the data. In Section 3, we give the theoretical explanation of our observations. In section 4, we describe the earlier theories. In section 5, we explain why our theory is closer to reality.

## 2. The data set and its analysis

Let us now have a look at some of our data. Figure 1 shows access statistics for three fixed content webpages from the website reverent.org (there are more examples in Simkin & Roychowdhury, 2008) which are apparently not affected by any decaying novelty factor. The absolute maximum of daily downloads happens two years after webpage publication (Fig.1 (a) and (b) ) and five years after publication (Fig.1 (c) ). The mentioned webpages are only two clicks away from site's front page. This assures them decent placing in internet search. For example, during the year 2010, 737 internet searches for "Donald Judd" had led to the webpage in Fig. 1(a) and 590 searches for "famous artist" had led to the webpage in Fig. 1 (c). This means that the webpages have some small but constant traffic, directed by search engines. Sometimes visitors mention the webpage, which they found in a blog, forum or social networking website. Sometimes a visitor to that blog reposts links in his blog, similarly to how scientific citations travel from one paper into another (Simkin & Roychowdhury, 2007). Sometimes this results in avalanches of blog entries. They lead to spikes of downloads, which are seen in Fig. 1. In the previous paper (Simkin & Roychowdhury, 2008) we had modeled these avalanches using theory of Branching Processes (Simkin & Roychowdhury, 2011). So we are not going to repeat the analysis in the present article. One reason why we put Fig. 1 in the article is that it helps to question the "novelty factor." Another reason is that it explains how we managed to get access statistics of web articles from many other websites.

Obviously, the number of referrals is proportional to the number of visitors to the referring webpage. Thus, when a blog links to one of the webpages of the website for which we have the access log, we can estimate the access statistics of that blog from the number of referrals. Here is a typical line from the access log:

```
67.188.206.95 - - [09/Feb/2008:12:03:46 -0700] "GET /an_artist_or_an_ape.html
HTTP/1.1" 200 3347 "http://www.metafilter.com/68935/Hey-my-Cheetah-could-paint-
that" "Mozilla/5.0 (Windows; U; Windows NT 5.1; en-US; rv:1.8.1.12) Gecko/20080201
Firefox/2.0.0.12"
```

At the beginning of the line is user's IP address. Next is the date and time of access. Afterward is the particular webpage, which was downloaded. 200 is the code that the download was successful. 3347 is the size in bytes. Next is referrer's URL. At the end is the browser information. We will need only three of these parameters for our research: date/time, downloaded webpage, and referring webpage.

In the above example, http://www.metafilter.com/68935/Hey-my-Cheetah-could-paint-that is the particular blog entry, which linked to the webpage http://reverent.org/an_artist_or_an_ape.html . However, while the blog entry is fresh it appears on blog's frontpage. So one does not have to visit the specific webpage containing the separate blog entry, but can be referred from blog's frontpage. Thus, the majority of referrers look like `http://www.metafilter.com/` . The frontpage is not the only additional referrer. There are also referrals from the next pages (for example, `http://www.metafilter.com/index.cfm?page=8`) or from tags (for example, `http://www.metafilter.com/tags/art` ). Therefore, to get all the referrals one has to select (for example, using "grep" command) all the lines in the access log files that contain domain name "metafilter.com." This, however, creates another complication. Metafilter.com has in addition the entry `http://www.metafilter.com/60584/My-mother-is-a-fish` which links to another webpage from the same site, `http://reverent.org/sounds_like_faulkner.html` . Therefore we have to select from all the access log lines that contain "metafilter.com" the lines that in addition contain "an_artist_or_an_ape.html". These will be all the referrals to `http://reverent.org/an_artist_or_an_ape.html` generated by the `http://www.metafilter.com/68935/Hey-my-Cheetah-could-paint-that` blog entry.

Figure 2 shows the number of referrals as a function of time for four different blogs. We use not calendar days to plot the data, but 24-hr days since first referrals. Thus if the first referral occurred at time *t* on day *d*, the first 24-hr day includes referrals up to time *t* on day *d*+1 and so on. We computed the average number of daily referrals using logarithmic binning with base 2. The error bars show the 95% confidence interval. We computed them using the number of referrals in the bin and assuming that referrals follow a Poisson process. The assumption is probably not justified but at least it allows getting some estimate of errors. We selected for this figure the blogs that linked to two or three different webpages from reverent.org. These were two or three separate entries, not links to several webpages in a single blog entry. Different symbols in Fig. 2 refer to different webpages. Lines are linear fits of these log-log plots and the parameters of fits are next to the lines. One can see a power law decay of the number of referrals as a function of time since link publication. One can also see that the exponents of the power law, though very different for different blogs, are very close for different entries of the same blog.

Figure 3(a) shows the histogram of the power-law exponents for 151 blog entries and news articles from 111 different websites. The data includes 8 different blog entries from 5 different bloggers on Blogspot.com, 12 blog entries from 10 different bloggers on livejournal.com, numerous blogs housed by lesser known sites, like livedoor.jp, a dozen news articles on lesser known news sites like thestranger.com, and a several articles in well known journals like economist.com. The total number of referrals in the dataset ranges from 38,512 (this came from ayacnews2nd.com ) to 109. In Figure 3(b) we give the Zipf's plot of the number of referrals. One can notice that the plot is fairly close to a straight line. So we have another power law – in the distribution of the number of referrals. This law is related to Zipf' law in the number of webpage downloads observed by Cunha, Bestavros, & Crovella (1995) and Nielsen (1997).

To obtain the data for Figure 3 we studied all blog entries and news articles (there was a restriction that the referrals should not be from forums for the reason we will discuss later) linking to reverent.org or ecclesiastes911.net websites, which produced over 100 referrals. There were 188 of suitable articles and 151 of them had shown a power law and 37 did not.

## 3. Our theory of the power-law decay of attention

We propose the following explanation. The probability that visitors to the webpage follow certain link, posted on this webpage, depends on the position of the link. They follow the current top link with highest probability. The second link they follow with smaller probability than the first and the tens with even smaller probability. One would not expect that position changing from first to second would have the same effect as changing from tenth to eleventh. It would be more natural to assume that proportional decrease in position results in proportional decrease in access probability. So that falling from first to second place reduces access probability by the same factor as falling from tens place to twenties. Mathematically this is expressed as

$$\frac{\Delta n(r)}{n(r)} = -\beta \frac{\Delta r}{r}. \qquad (2)$$

Here $n(r)$ is the access rate and $r$ is the position rank (so that the top link has rank 1, the second – 2 and so on) and $\beta$ – some proportionality coefficient. This results in a power law decay of access probability with link's position:

$$n(r) \sim 1/r^{\beta}. \qquad (3)$$

We cannot determine the value of the exponent $\beta$ by this reasoning. To do this we need to say by how much probability of access decreases when the position falls from first to second. For example, if it decreases twice, then the exponent will be exactly 1.

How is this related to time-dependence of access probability? The simplest case is the "Reality Carnival" site (see Fig. 2(a)). This is a webpage containing links to other webpages, which the owner, Dr. Pickover, found interesting. There is no separate blog entry for each link – just the link and a few word description. Dr. Pickover adds one link a day. The new link goes on the top of the list and all already present links fall in their position by 1. A year worth of links is on the page at a time. Thus, the number of days passed since link addition exactly corresponds to link's position in the list. Therefore, $r = t$, where $t$ is time passed since link publishing, measured in days. We thus get Eq. (1).

The probability of following a link depends not only on its position in the list, but also on how attractive is link's description. The attractiveness factor is constant and does not vary from day to day. Naturally, it influences only a prefactor and not the power law exponent. At different times, Reality Carnival linked to two webpages from reverent.org. One can see from Figure 2(a) that the prefactors differ by a factor of 1.5, but the exponents are the same within 1%. Similar pattern holds for three other blogs shown in Figure 2.

One may argue that our derivation has a flaw since it implicitly assumes that users' clicking behavior is independent of their previous visits to the website. In reality, it is likely that the user is not going to follow the link, which he had followed during the previous visit. At the same time Eq.(3) gives a non-zero probability for such event. Fortunately it is enough to assume that each user reads only a small fraction of news stories or blog entries appearing on any given news site or blog. Suppose that this fraction is 10%. And let us assume that the user never reads the same story twice. Let us also assume that the number of people who will look at the link to the story is given by the power law: $n_l(t) \sim 1/t^{\beta}$. Suppose that the fraction of the users who have read the article of age $t$ is $f(t)$. The effect of not reading the same thing twice will be that the number of the users who will follow the link will be $n(t) = (1 - f(t))n_l(t) \sim (1 - f(t))/t^{\beta}$. Obviously $f(0) = 0$ and, in our example, $f(\infty) = 0.1$. Thus, the power law will be multiplied by a function of time, which falls from 1 to 0.9, as the story grows older. We are not going to notice this on a logarithmic scale.

Other blogs are different from Reality Carnival: they may add several new items a day, or may add only one item in several days. Great majority of them move earlier entries to next pages. In such case, we can plot the number of referrals as a function of page number. Figure 4 shows such data. The number of referrals falls as a power of the page number. This gives more credence to our claim that the probability of following a link is determined by its position on the website.

## 4. Earlier theories

Dezsö et al (2006) had analyzed a month of access log of a Hungarian news website. They reported that the average rate of accessing a news story falls as a power law of the time passed since its publication (see Eq. (1)) with $\beta = 0.3 \pm 0.1$. They proposed the following theoretical explanation. They found that the distribution of time-intervals $\tau$ between the visits by the same visitor follows a power law $p(\tau) \sim 1/\tau^{\alpha}$ with $\alpha = 1.2 \pm 0.1$. They speculated that visitors access all news items of interest to them that appear on the website since their last visit. The number of visitors who did not yet see the document of age $t$ is $n(t) \sim \int_t^\infty 1/\tau^\alpha \sim 1/t^{\alpha-1}$, which means that $\beta = \alpha - 1$. The observed values of $\alpha$ and $\beta$ agree with this theory. There are, however, problems with the above explanation. First, it is not clear from the article what are visitors and what are times between visits. There are two ways of tracing visitors: through cookies and through IP addresses. There is more uncertainty in the definition of a visit. For instance, popular web analytic tool AWStats (2010) separates visits when there was over an hour between requests from the same IP. This is quite arbitrary. Therefore, we had to ask the authors what they meant by their words. It turned out (Dezsö, 2009) that the visitors were determined from IPs and "visits" were all HTML requests, that is, every line in the access log. This includes not only webpages but also all image files. Thus downloading one webpage with several images produces a number of "visits" with several intervals between them. That is why the power law in Fig. 4(a) of Dezsö et al (2006) spreads into the region of one-second intervals between "visits." One still could separate real visits: those with interval of a day or more can without doubts called different visits. However, Dezsö et al (2006) did no analysis showing that after a long absence a visitor looks at more webpages, than after a short absence.

Another concern is that many different users have same IP: this is certainly a problem studying access log of a major news website of a small country. This problem, however, would not affect the work of Chmiel, Kowalska, & Hołyst (2009) who investigated the access logs of two Polish portals. They separated different users not by IP addresses, but by using cookies. Nonetheless, their Figure 4 is very similar to Fig. 4(a) of Dezsö et al (2006). The exponent they report is about 1.3 which is very close to the 1.2 value reported by Dezsö et al. Note however, that the Figure 4 of Chmiel, Kowalska, & Hołyst (2009) which shows the same type of data as Fig. 4(a) of Dezsö et al (2006) has a different caption. While the figure from Dezsö et al has the caption "The distribution of time intervals between two consecutive visits of a given user" the corresponding figure from is called "The distribution of time spent by the user on one subpage" which is, of course, a far more reasonable interpretation when we talk about few minutes intervals.

Another problem arises if we try to apply the theory Dezsö et al (2006) to our data. Some of the referrers show power law decay for several years (see Figures 3 and 4). It would be strange to suspect that some users visit with several years' intervals and then look up everything they missed.

Wu and Huberman (2007) had studied the time series of the number of "diggs" (bookmarks) on digg.com website. They proposed the following model. The evolution of the cumulative number of diggs, $N(t)$, is described by the equation $\frac{dN(t)}{dt} = N(t)r(t)$ where $r(t)$ is a decay factor. The equation has the solution $N(t) = N(0)\exp\left(\int_0^t dt' r(t')\right)$. And the number of new diggs is equal to $n(t) = \frac{dN(t)}{dt} = N(0)\exp\left(\int_0^t dt' r(t')\right) r(t)$. In the limit of large $t$ this becomes $n(t) = N(0)\exp\left(\int_0^\infty dt' r(t')\right) r(t) \sim r(t)$. They got the best fit to the actual data using the decay factor $r(t) = \exp(-0.4 t^{0.4})$. Note that a stretched exponential looks very similar to a power law. By looking at Figure 3 of Wu and Huberman (2007), one can guess that $r(t) = 1/t^{1.5}$ would do almost as good. This agrees well with our data.

Hogg & Lerman (2009), who also studied digg.com, explained decay of attention using visibility factor instead of novelty factor. The story first appears on website's front page, as time goes, newly added stories push it down to the page two, page three and so on. The further the story is from the front page the less visible it is, less people notice it, and less new "diggs" it generates. When new stories are added at a constant rate the position of the story on the website is proportional to the time passed since its publication. The frequency of accessing the story is thus some decreasing function of time. To find the exact functional dependence Hogg & Lerman (2009) used the model introduced by Huberman et al (1998) to describe the probability distribution of the number of webpages a user views on a certain website.

The model assumes that each page, *i*, that the user visits has a certain value $V_i$ for this user. In addition it assumes that the value of the next page the user visits is stochastically related to the value of the current page: $V_{i+1} = V_i + \varepsilon_i$, where $\varepsilon_i$ are independent and identically distributed Gaussian random variables. Thus the value performs a Brownian motion. The user continues to browse the website until the value becomes negative. The distribution of the number of viewed webpages, *k*, is thus the distribution of the first passage times. This distribution is inverse Gaussian. In the case of zero drift it has a large *k* asymptotic

$$p(k) \sim k^{-3/2}.  \qquad (4)$$

Actually, if instead of a Gaussian distribution we will take a distribution where $\varepsilon_i$ is +1 or −1, then we will get an ordinary random walk and the last result will be a bit easier to derive (see Feller, 1957).

Hogg & Lerman assumed that the user starts browsing digg.com starting with the front page and afterwards proceeds to page 2, page 3 and so on. Thus to the page *k* will get the users who visited *k* or more pages of the website. This will be a cumulative distribution of Eq. (4): $p(k) \sim k^{-1/2}$. Since the page number the story is on, is proportional to time passed since story publication, the theory of Hogg & Lerman leads to Eq.(1) with $\beta = 1/2$. This appears to be at odds with what we see the Figure 3 of Wu and Huberman (2007) which is consistent with $\beta = 1.5$.

## 5. Discussion

Although the functional form of the frequency of access decay function reported by Dezsö et al (2006) agrees with our data, the value of the exponent does not. They reported $\beta = 0.3$, while we always get $\beta > 0.6$ and in 98% of the cases $\beta > 0.8$. Why did Dezsö et al (2006) observe $\beta = 0.3$? Earlier we mentioned that 37 of 188 studied blogs and news articles did not show power-law decay. Figure 7 shows a typical example. This is an article in Significance magazine. From Figure 7(a) we see that there is no power-law fit. And from Figure 7 (b) we can see why: apart from the initial peak at the release of the article, there are several other peaks. The reason is that the article was discussed in several blogs and forums. The visitors were coming to the article from these blogs and forums and the position of the article on the website of Significance magazine was irrelevant. Dezsö et al (2006) show in their Figure 3 only average data (over almost four thousand news items). We suspect that if the data for separate news stories were available the plots would look for some news item similar to Figure 7 of the present manuscript. The power law with $\beta = 0.3$ could be an artifact of averaging over many news items with different access patterns.

In a related study, Leskovec et al (2007) reported that the average number of new in-links to a blog entry falls of as an inverse power of time passed since publication of this blog entry. The exponent of the power law is 1.5. We may reason that the number of new in-links is proportional to the number of views (just like the number of referrals). The findings of Leskovec et al (2007) agree with our results (see Figure 3(a) of that paper).

In another related work Johansen & Sornette (2000) had studied how the number of downloads of their paper behaved after the url was published in a newspaper. They found that after initial peak the number of downloads fell of as a power law with the exponent of 0.6. We indeed had once seen such a shallow exponent; however, Johansen and Sornette looked at a bit different thing than what we have looked at. They studied the total number of downloads, not the number of referrals from newspaper. It is certain that newspaper article triggered a cascade of blog entries linking to the paper. Thus, exponent 0.6 describes not the decay of accessing the newspaper article, but the effect of the whole cascade. We can illustrate this with our data. In Fig. 1(a), you can see a peak in October 2009. It is a result of a cascade triggered by the publication of the link in the popular blog boingboing.net. Figure 6 shows the number of total downloads and the number of referrals from boingboing.net. While the number of downloads relaxes as a power law with exponent 0.9, the number of referrals from boingboing.net falls off with the exponent 1.5.

The idea of novelty factor is very reasonable, and the effects talked about by Dezsö et al (2006) may also take place. However these theories cannot account for all of the experimental observations.

Some evidence in support of the importance of the visibility factor comes from the study of forums. The forums differ from blogs and news sites in how comments affect the position of the story. In the case of blogs and news articles, the comments do not change their position on the website. In contrast, when somebody adds a new reply to the forum thread, the thread goes to the top of the forum. Figure 5 shows referrals from two forums together with the number of posts added to the thread on the given day. One can see that each spike in referrals is accompanied by at least one new post. This means that some forum user resurrects the old thread by posting into it and raising it to the top of the forum. Interestingly for the Figure 5(b) the highest peak of referrals happens two month after the thread was started. If one would attempt to use the "novelty factor" of Wu and Huberman (2007) to explain this, one would have to assume that the novelty increased as time passed. Dezsö et al (2006) theory will not account for this as well.

We did an experiment to explicitly test the effect of link position on clicking probability. We took one already existing webpage, which contained a list of 23 links. For the purpose of this study we added to the webpage a PHP script, which rotated these 23 links on daily basis. So that the next day the second link becomes the first, the third becomes second, and the first becomes 23rd. The rotation was implemented to average out the effect of different links being not equally attractive to the visitors. The data shown in Figure 8 were collected over four months. The data clearly show a decay of the number of clicks with link's position, which can be approximated by a power law with $\beta = 0.4$. This shows that the position effect obviously exists. However the value $\beta = 0.4$ is much less than the typical $\beta = 1.5$ (see Figure 3) which describes the decay of attention to news articles with time. So the position effect is probably not the sole reason of the decay of attention. Probably this decay is the result of a combination of several factors including those suggested in earlier theories.

This work was supported in part by the NIH grant No. R01 GM105033-01.

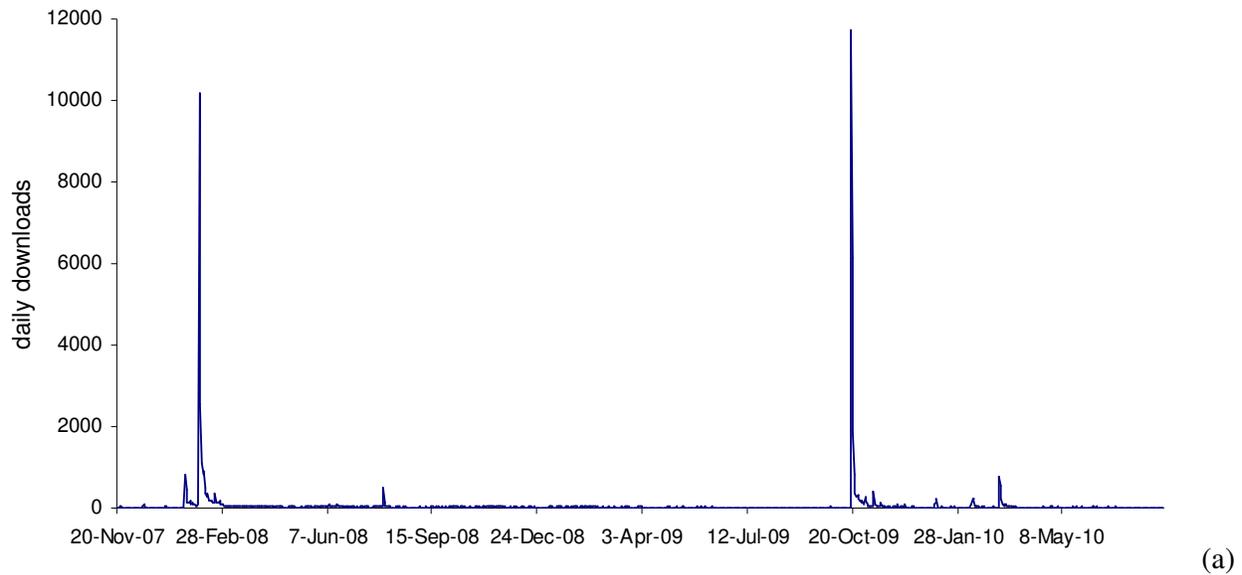

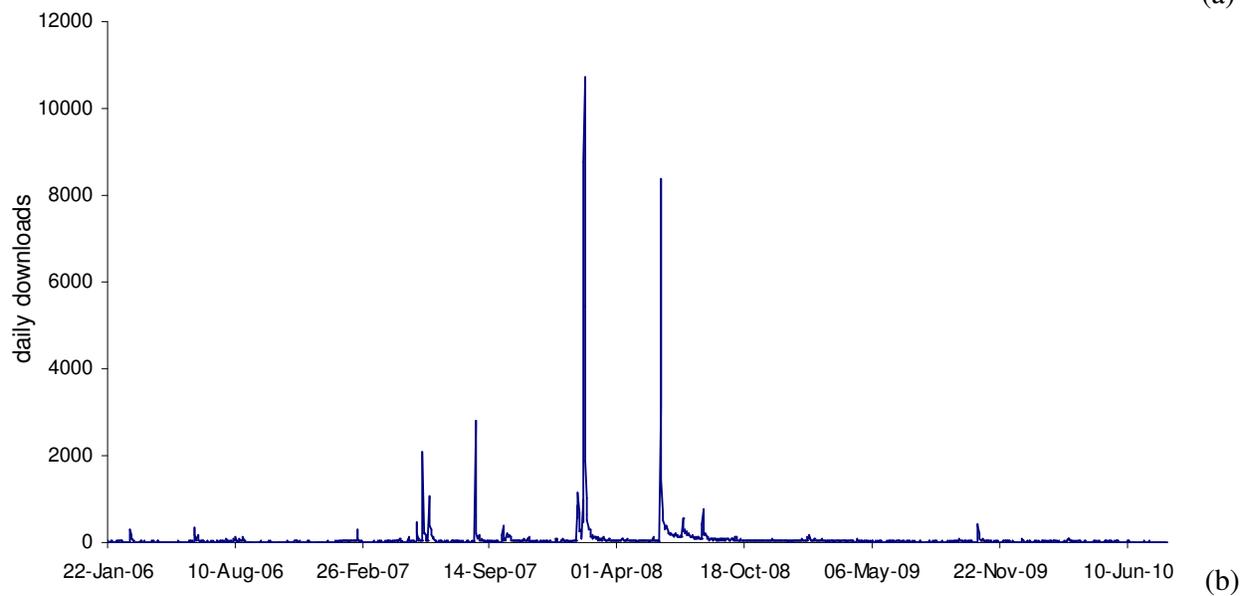

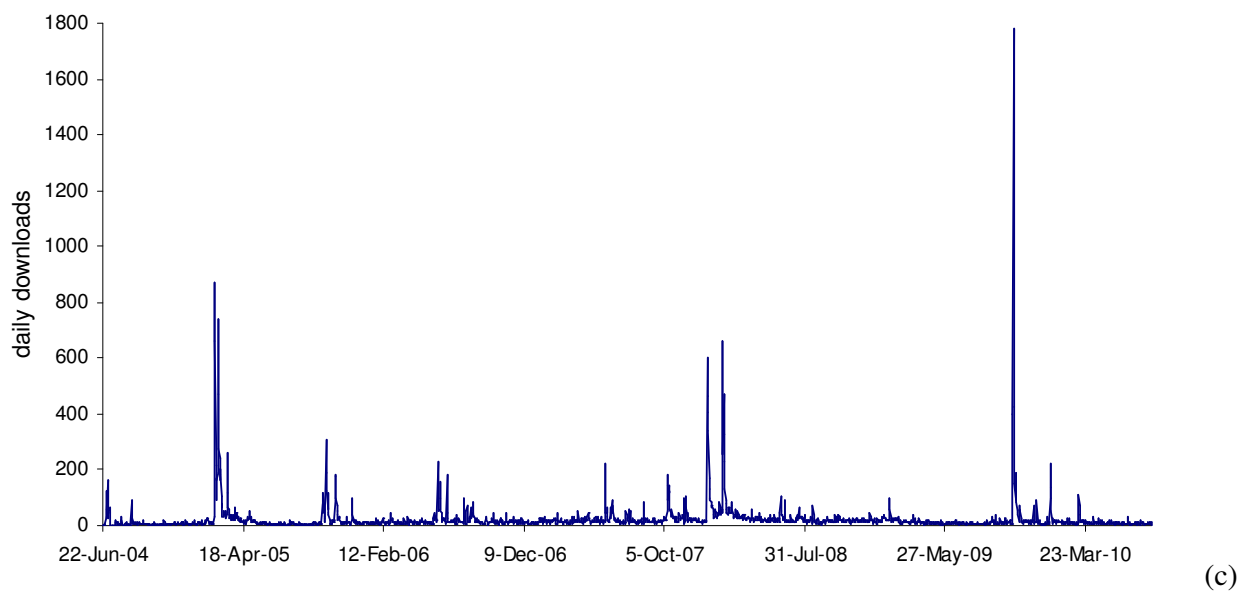

**Figure 1.** Access statistics for three webpages: (a) http://reverent.org/donald_judd_or_cheap_furniture.html , (b) http://reverent.org/an_artist_or_an_ape.html , and (c) http://reverent.org/great_art_or_not.html since the day of their release and until 8/12/2010.

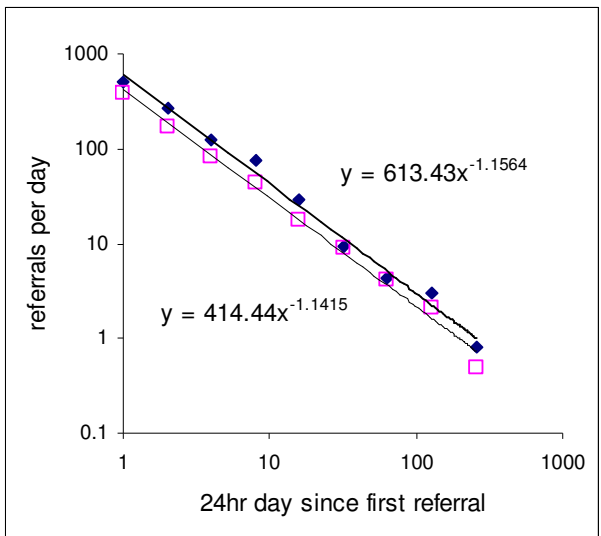
(a)

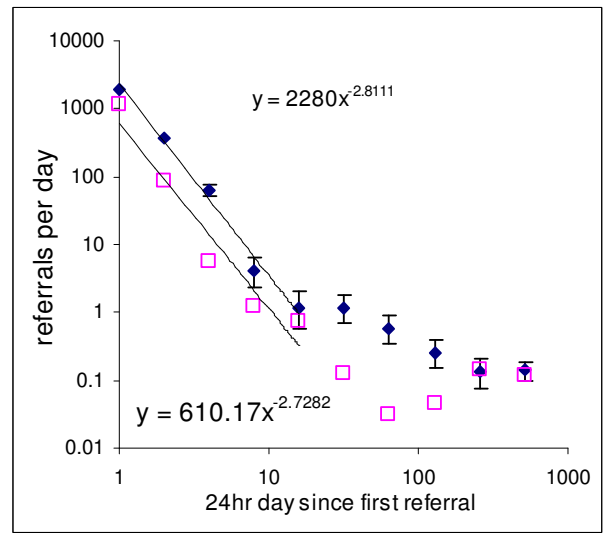
(b)

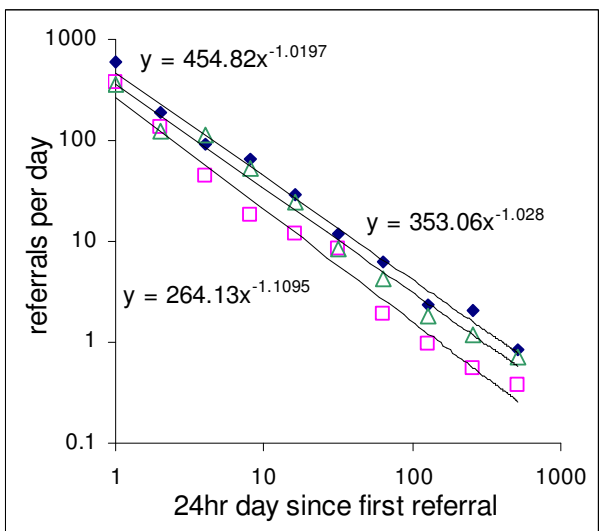
(c)

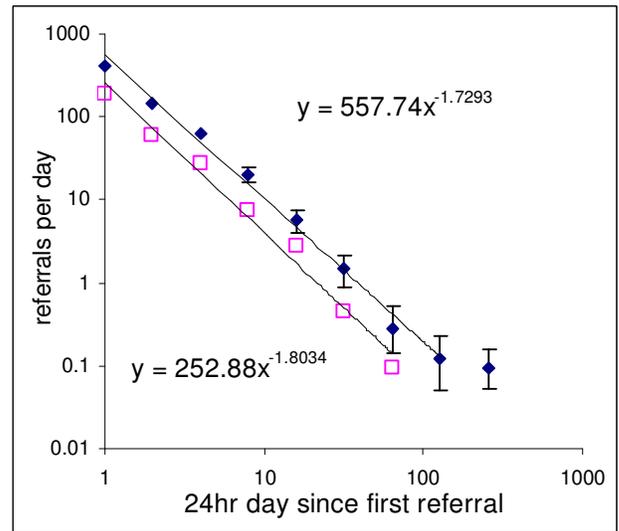
(d)

**Figure 2.** (a) Referrals from **http://sprott.physics.wisc.edu/pickover/pc/realitycarnival.html** . It at different times linked to two webpages from reverent.org. One of them is shown by solid rhombs and another by empty squares. The lines are the least-square fits to a power law. Although pre-factors are 50% different, the exponents differ only 2%.(b) Referrals from **http://www.metafilter.com/** . (c) Referrals from **http://howaboutorange.blogspot.com** , which linked to three different pages from reverent.org. Note that it was not a single post linking to three webpages, but three different posts widely separated in time. (d) Referrals from **http://krylov.livejournal.com/** . We show errors for one data series (solid rhombs) in each plot in the case when those errors are bigger than symbol size. We do not show errors for other data series to avoid mess.

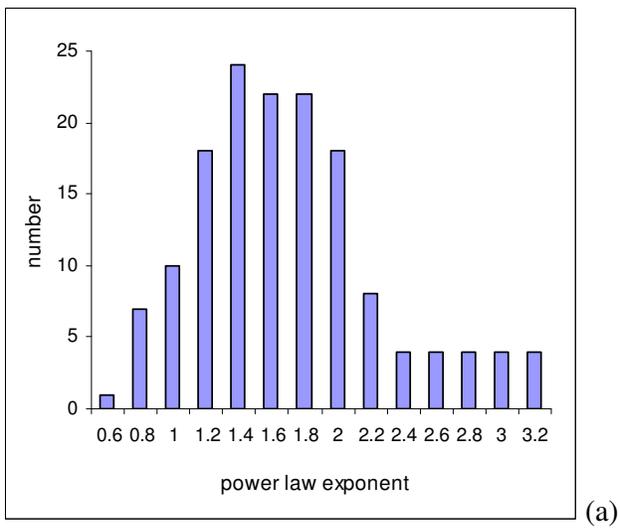

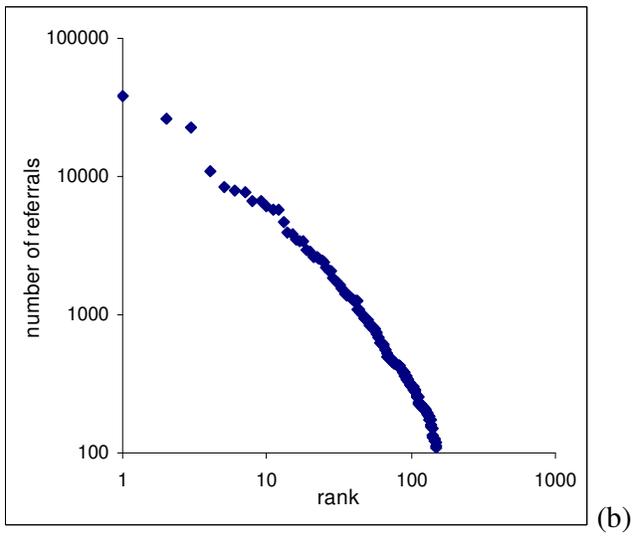

**Figure 3.** (a) The histogram of the power-law exponents for 151 blogs / news articles. The bin is 0.2. (b) The distribution of the number of referrals.

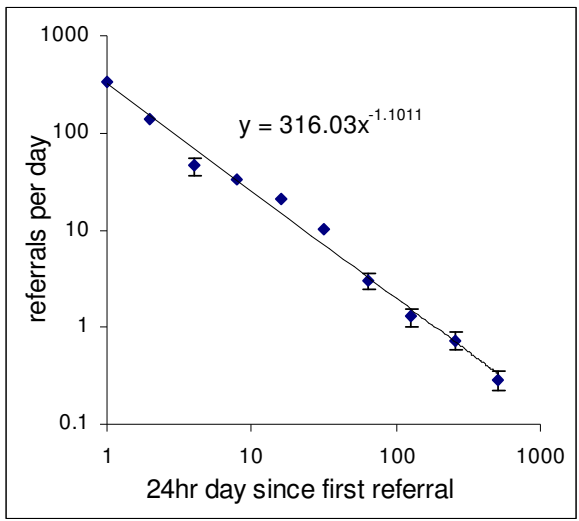
(a)

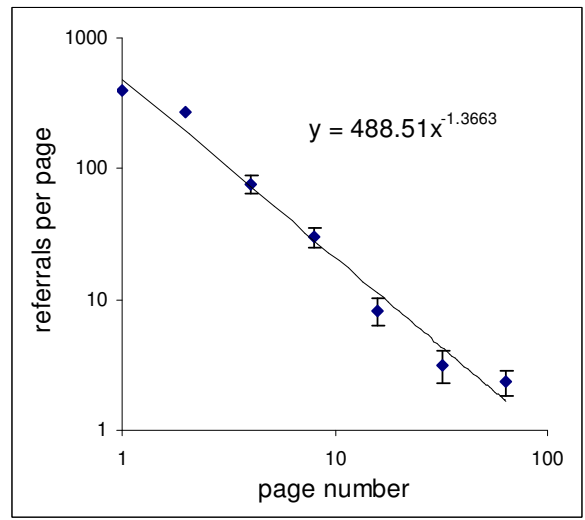
(b)

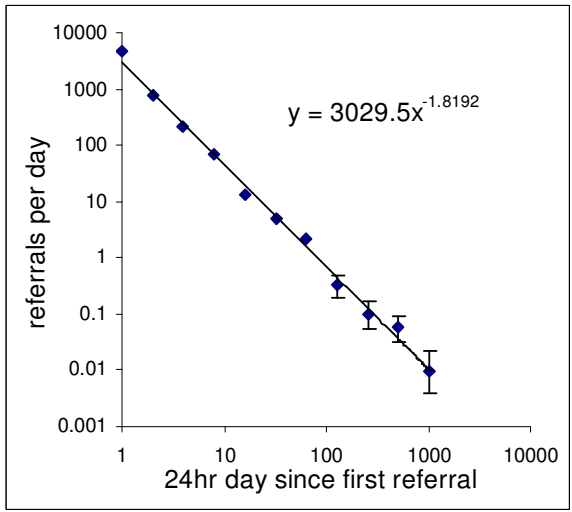
(c)

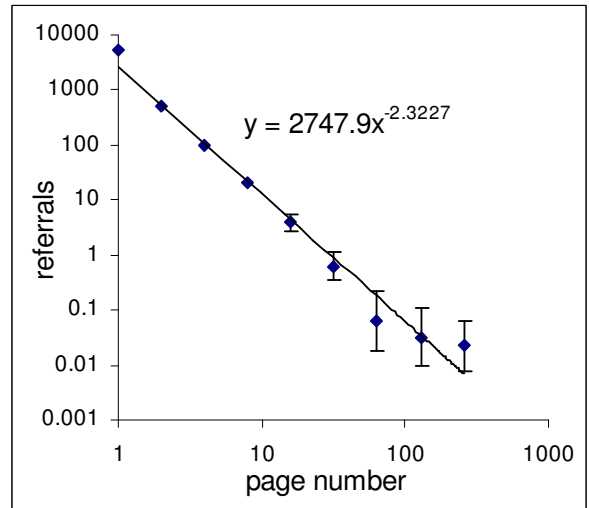
(d)

**Figure 4.** (a) Referrals from http://www.museumofhoaxes.com. (b) The distribution of the referrals by page number. Distribution of referrals from http://www.flabber.nl/ (c) by day (d) by page number.

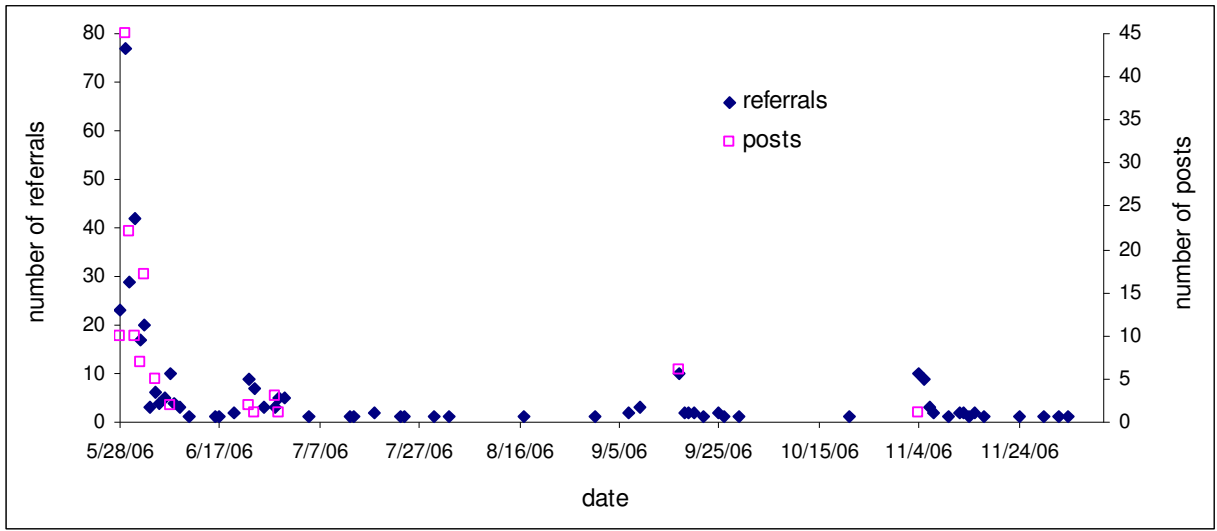

(a)

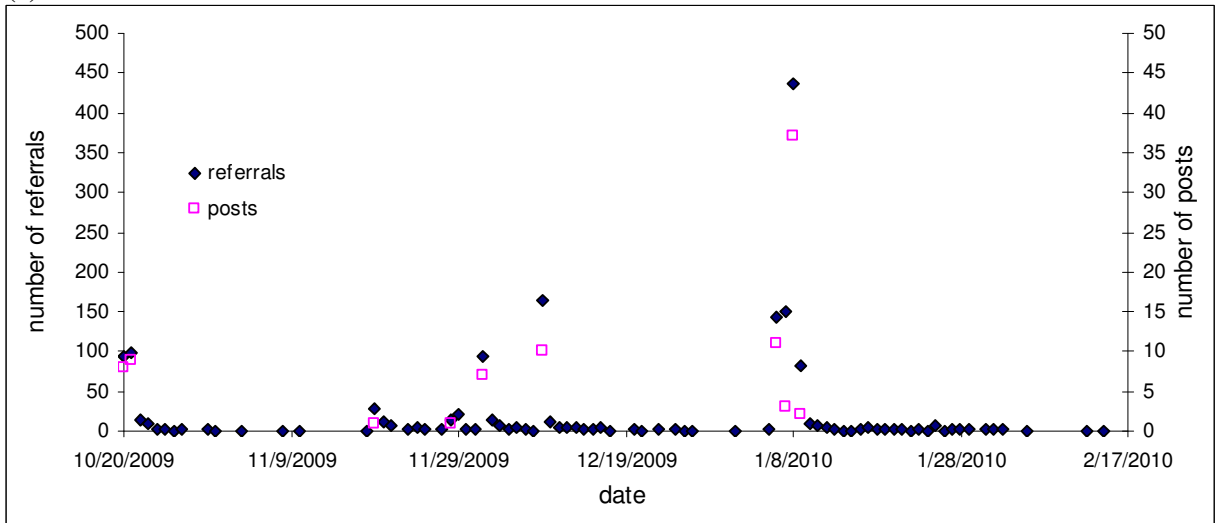

(b)

**Figure 5.** (a) Daily referrals to **http://reverent.org/ru/quizzes.html** from **http://forum.ixbt.com/topic.cgi?id=65:1292** shown alongside with the daily number of posts in the thread. (b) The same for referrals to **http://reverent.org/true_art_or_fake_art.html** from **http://www.douban.com/group/topic/8380611/** . We did not plot zero numbers in the figure. So if you see a rhomb very close to a horizontal axis – it corresponds to at least one referral.

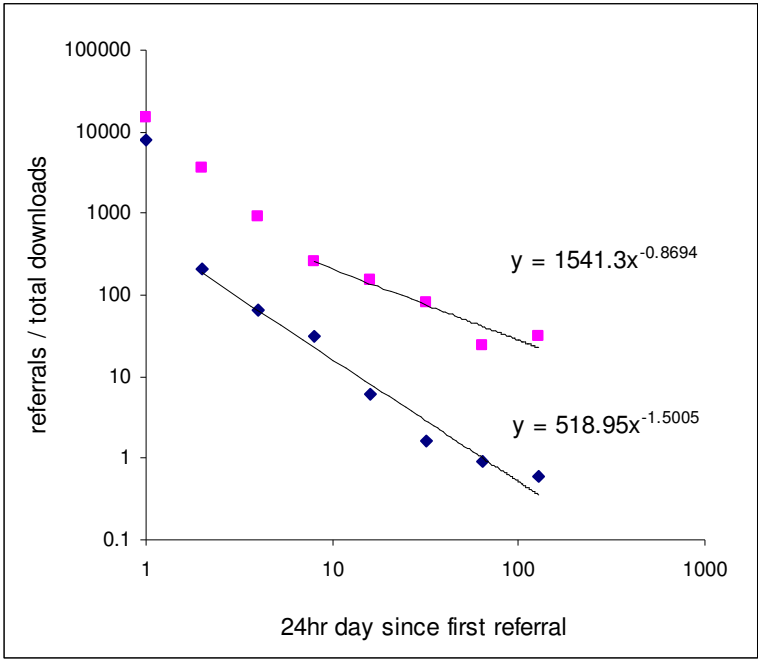

**Figure 6**. The rhombs show referrals from boingboing.net, which linked on 10/19/2009 to the webpage whose access statistics is given in Fig. 1a. The squares show the total number of downloads of the webpage in question.

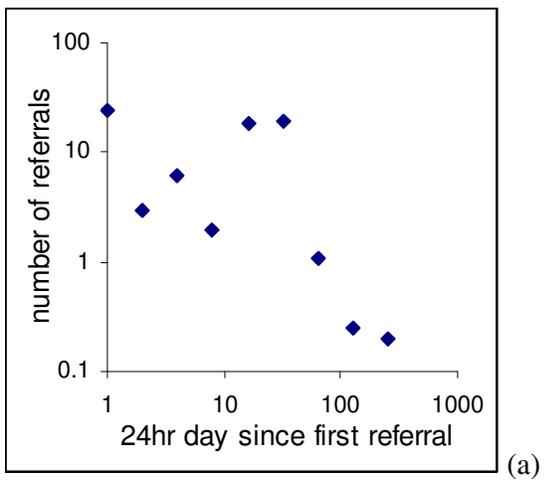

(a)

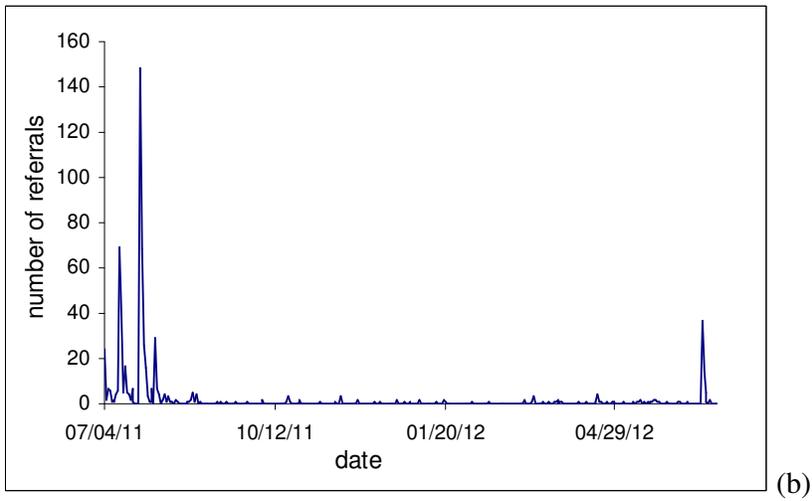

(b)

**Figure 7.** Referrals from http://www.significancemagazine.org/details/webexclusive/1237447/PhDs-couldnt-tell-an-actor-from-a-renowned-scientist.html (a) In log-log coordinates. (b) In linear coordinates.

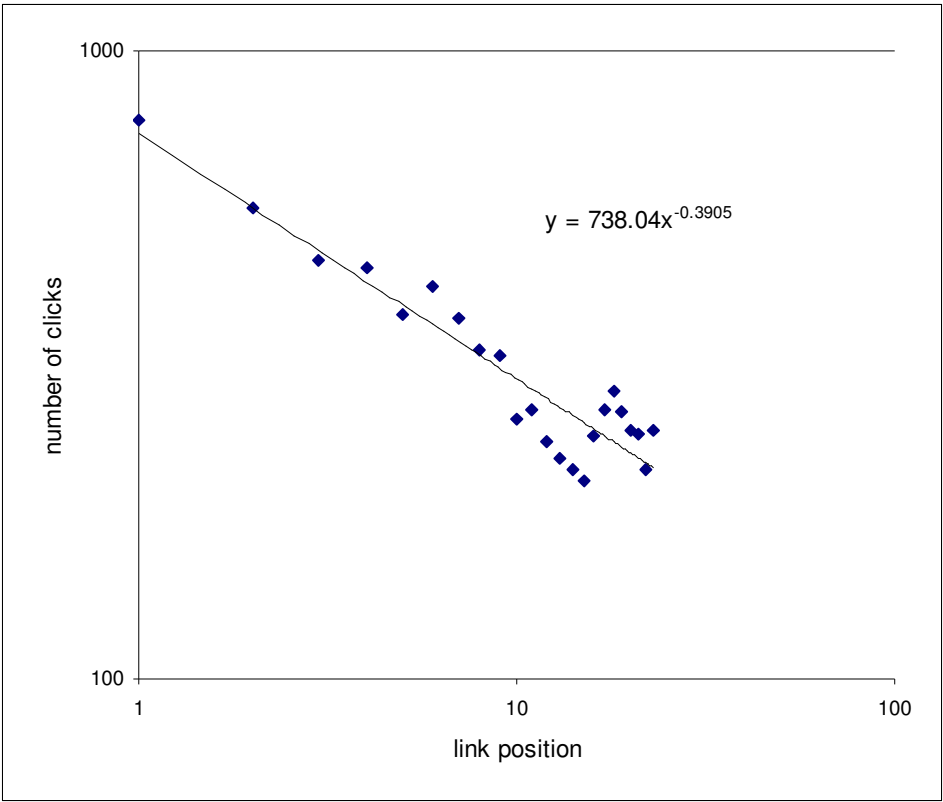

**Figure 8.** Number of clicks versus link position. The webpage's 23 links were rotated on daily basis for 4 months.